\documentstyle[aps,epsfig]{revtex}
\begin{document}
\def\H{{\cal H}}
\def\S{{\cal S}}  
\def\L{{\cal L}}
\def\O{{\cal O}}
\def\simlt{\stackrel{<}{{}_\sim}}
\def\simgt{\stackrel{>}{{}_\sim}}

\title{Moduli and Monopoles}

\author{Ram Brustein and Irit Maor}

\address{Department of Physics, Ben-Gurion University, 
Beer-Sheva 84105, Israel\\
email: ramyb,irrit@bgumail.bgu.ac.il}

\maketitle{
\begin{abstract}

Long periods of coherent oscillations of moduli fields relax the bound on
possible initial monopole density in the early universe, and in some cases
eliminate it completely.

\end{abstract}}

\ \\
\centerline{Preprint Number: BGU-PH-98/14}
\pacs{PACS numbers:  98.80.Cq,14.80.Hv}

\section{Introduction}

String theory  contains moduli fields -- 
massless fields that move on string ground state
 manifolds \cite{polchinski}.  If supersymmetry is unbroken,  these
 massless fields remain massless to all orders in perturbation theory, 
but it is assumed that all moduli obtain mass through
non-perturbative interactions at some high scale, with perhaps a
few exceptions. Among the many moduli the dilaton is  particularly 
interesting because it determines the string and gauge coupling. 
String moduli have only nonrenormalizable couplings to light fields 
and their typical range of variation is the Planck scale.

Monopoles are one-dimensional topological defects that carry a magnetic charge,
but their main relevant attribute is that
 they behave as stable non-relativistic (NR) particles 
 (see, for example, \cite{linde,eu}). In addition to the good
old GUT monopoles, there are many  stringy monopoles, dyons and other
exotic creatures \cite{polchinski}. Since some grand
symmetry is expected to brake  into a lower one, monopoles and other exotics
are  expected to be  produced via the Kibble mechanism (see however
\cite{denial1,denial2,denial3,denial4}).  GUT type monopoles are the most dangerous objects, since they
are expected  to have  masses of the order of $M_m\sim 10^{16}~GeV$, and an
initial abundance  of $(\rho_m/\rho_r)_c\approx 10^{-11} (M_{m}/10^{16}
GeV)(10^{14} GeV/T_c)$, where $\rho_m$ is the monopole energy density,  $\rho_r$
is the radiation energy density, and $T_c$ is the temperature when monopoles were
created \cite{eu}.

Because of their NR nature, the energy of monopoles decreases at a slower rate 
than that of radiation, leading, if left alone, 
to early monopole domination \cite{pres}. The
Kibble mechanism predicts a present abundance of $\rho_m\approx 10^{11}\rho_c$,
where $\rho_c$ is the critical energy density,
But, observational limits on the presence of monopoles today,
imply that the fraction of monopole to the critical energy density does not
exceed unity $\rho_m/\rho_c<1$. This discrepancy is referred to as the monopole  
problem. Any other NR relics produced early enough will present the same
difficulties, and since we will use only  NR nature of monopoles for our analysis,
 our results are applicable to other NR relics as well.

One class of proposed solutions argues that monopoles are 
produced in  lower abundances or go through a phase of 
annihilation \cite{lp,denial1,denial2,denial3,denial4}.
Another type of solutions is based on additional non-adiabatic expansion so that
$\rho_m/\rho_c$ gets diluted. In order to reconcile the theoretical and
observational limits, the equivalent of 27  efolds of volume 
expansion should be supplied.
Inflationary models, assuming inflation does occur after monopoles were
produced, can supply much more than the needed 27 efolds 
\cite{inflation,linde,eu}. \\

We set out to explore the possible influence of moduli on the monopole
problem.\\

We show here that
long periods of coherent oscillations of moduli   
 can replace enough inflationary expansion and may relax  the monopole
problem.  In general, moduli start out displaced by about a Planck distance
 from the global
zero-temperature minimum of their potential, so when the universe cools down,
they start to coherently roll or oscillate,
creating particles as they do. During periods of coherent oscillations the
universe  is effectively matter dominated (MD), and the relative  growth of
monopole density  slows down, therefore  the  bound on allowed initial density
of the monopoles relaxes. If the duration of the coherent oscillation epoch is
long enough, the bound is eliminated altogether.

If moduli energy density is low and their mass lower than the Hubble expansion
rate, they are essentially frozen, the only
possible exception being the dilaton, since it develops a potential due to the
existence of monopoles 
\cite{Tseytlin:1992xk,Barreiro:1998aj,banks2,Choi:1998dw,Choi:1998nx}. 
However, it turns out, as we will show,
 that the change from standard radiation dominated (RD)
cosmology is small. The interesting situation is when
 moduli density is high, and deviations from standard cosmology are large. 

A similar idea has already been discusses in the context of axions, and in
other work on modular cosmology,  \cite{Lazarides:1990xp,Banks:1996ea}.

The effective equations of motion in a cosmological background with 
a massive  dilaton included are well known,
   \begin{eqnarray}
 & 3H^2=\frac{1}{4}\dot{\phi^2}+\frac{1}{4}m^2\phi^2+
 \frac{1}{2}e^{2\phi}\rho &
\label{ein1} \\
         &
2\dot{H}+3H^2+\frac{1}{4}\dot{\phi}^2-\frac{1}{4}m^2\phi^2+
\frac{1}{2}e^{2\phi}p=0
& \label{ein2} 
\\
      & \ddot{\phi}+3H\dot{\phi}+m^2\phi+\frac{1}{2}e^{2\phi}(\rho-3p)=0 &
\label{dil} \\
& \dot\rho+3(H+\frac{1}{2}\dot\phi)(\rho+p)=0, & \label{cons}  
 \end{eqnarray}
where we are using units in which  $m_p\equiv\sqrt{16\pi}$, where $m_p$ is
the Planck mass.
 The  conservation equation (\ref{cons}) for any
additional radiation or matter is not independent of the other equations, but we
include it for completeness.
Looking at (\ref{dil}), the dilaton couples to NR matter but not 
to radiation, since for radiation $\rho-3p=0 $. In the 
case of a universe with radiation as the only source, $ \phi=Const.$ 
is a solution and we retrieve the standard Friedmann-Robertson-Walker 
(FRW) cosmology.

For completeness we also include the standard equations for any of the other
moduli fields, assuming the dilaton is constant and fixed at the correct expectation
value, 
\begin{eqnarray}
 & 3H^2=\frac{1}{4}\dot{\phi^2}+\frac{1}{4}m^2\phi^2+
 \frac{1}{2}\rho &  \label{mod1}\\
         &
2\dot{H}+3H^2+\frac{1}{4}\dot{\phi}^2-\frac{1}{4}m^2\phi^2+
\frac{1}{2}p=0
& \label{mod2}\\
      & \ddot{\phi}+3H\dot{\phi}+m^2\phi=0 & \label{mod3}\\
& \dot\rho+3 H  (\rho+p)=0, &  \label{mod4}
 \end{eqnarray}
The qualitative behaviour of solutions to (\ref{mod1}-\ref{mod4})
are well known,  for high values of $H$ the field is frozen, 
and then it  starts to oscillate when $H$ decreases below its mass $m$.

In section II we show that for low moduli and dilaton density there are no
substantial deviations from standard RD cosmology, in section III 
we analyze the case of high moduli density and in section IV we 
discuss our results and their validity.

\section{Low Moduli Density}

As already noted, the only possibly interesting field among moduli, for low
density is the dilaton, because the others are trivially frozen.
 The dilaton
density is given by  
\begin{eqnarray}
  & \rho_{\phi}=\frac{1}{2}\dot\phi^2+V(\phi) \label{dens} \nonumber \\
  & V(\phi)=\frac{1}{2}m^2\phi^2+\frac{1}{4}e^{2\phi}\rho_m. & 
\end{eqnarray}
We proceed to show that adding low dilaton density to a RD 
universe with a small amount of monopoles, does not substantially affect the 
standard evolution.  We will treat each contribution to the potential
separately using perturbation theory.

We begin with the potential due to NR matter. For a universe with radiation 
and a massless dilaton, equations (\ref{ein1}--\ref{cons}) take the form,
\begin{eqnarray}
  & 3H^2=\frac{1}{4}\dot\phi^2+\frac{1}{2}e^{2\phi}\rho_r &  \\
  & 2\dot H+3H^2+\frac{1}{4}\dot\phi^2+\frac{1}{6}e^{2\phi}
    \rho_r=0 &  \\
  & \ddot\phi+3H\dot\phi=0 &  \\
  & \dot\rho_r+4(H+\frac{1}{2}\dot\phi)\rho_r=0. & 
\end{eqnarray}
The solutions of these equations are well known,
\begin{eqnarray}
   & \phi=\phi_0=Const. & \label{radsol1} \\
   & \rho_r=6e^{-2\phi_0}H_0^2 & \label{radsol2} \\
   & \rho_r\propto \frac{1}{R^4},
~~R\propto \sqrt{t},~~H=\frac{1}{2t} & 
\end{eqnarray}
which is (up to scaling) standard RD cosmology. 

Now we will perturb the solutions, 
\begin{eqnarray}
  & \rho\rightarrow\rho_r+\delta\rho,
~~~\delta\rho\equiv\rho_m\ll\rho_r & \nonumber \\
  & H\rightarrow H_0+\delta H & \nonumber \\
  & \phi\rightarrow\phi_0+\delta\phi. & \nonumber 
\end{eqnarray}
Using (\ref{radsol1}--\ref{radsol2}), equations 
(\ref{ein1}--\ref{dil}) become,  to first order in the perturbations, 
\begin{eqnarray}
  & 6H_0\delta H=\frac{1}{2}e^{2\phi_0}\rho_m+6H_0^2\delta\phi &  \\
  & 2\delta \dot H+6H_0\delta H+2H_0^2\delta\phi=0 &  \\
  & \delta\ddot\phi+3H_0\delta\dot\phi+
  \frac{1}{2}e^{2\phi_0}\rho_m=0,& 
\end{eqnarray}
We assume that all terms in the equation are of the same magnitude, and 
that the solutions take a power dependence on time,
$ \rho_m\propto t^{\alpha},~~\delta H\propto t^{\beta},~~\delta\phi\propto 
    t^{\gamma}$, 
and we already know that $ H_0\propto t^{-1}.$
Demanding that all elements of each equation  have the same time dependence 
forces $\alpha=\beta-1=\gamma-2 $. Since, in addition,  we require
 $ \rho_m $ 
to describe monopoles in RD background,
$\alpha=-\frac{3}{2}$,
therefore:
\begin{eqnarray}
   & \rho_m\propto t^{-3/2},~~\delta H\propto t^{-1/2},~~\delta\phi\propto 
    t^{1/2} &. \nonumber
\end{eqnarray}
The graphs in Figure 1 show that the numerical solutions
of the exact equations indeed reveal such time dependence. 

\vbox{
\begin{figure} 
  \begin{center}
    
    \vspace{.5in}
    
    \epsfig{file=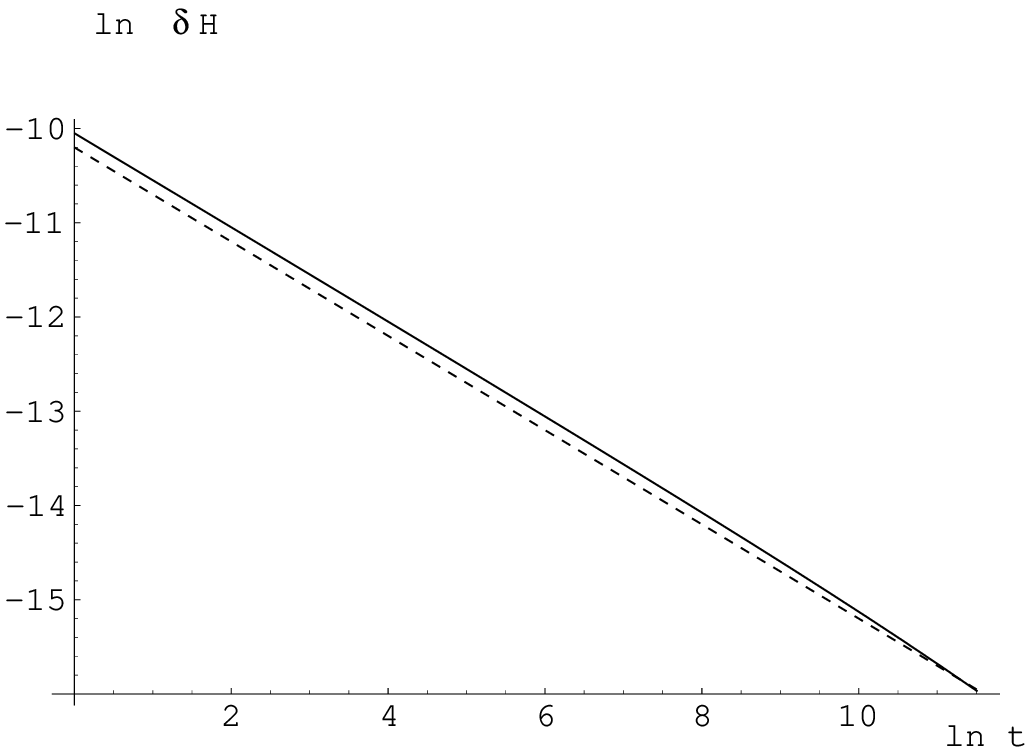,height=37mm,width=70mm} \ 
    
    \vspace{.5in}
    
    \hspace{-.3in}\epsfig{file=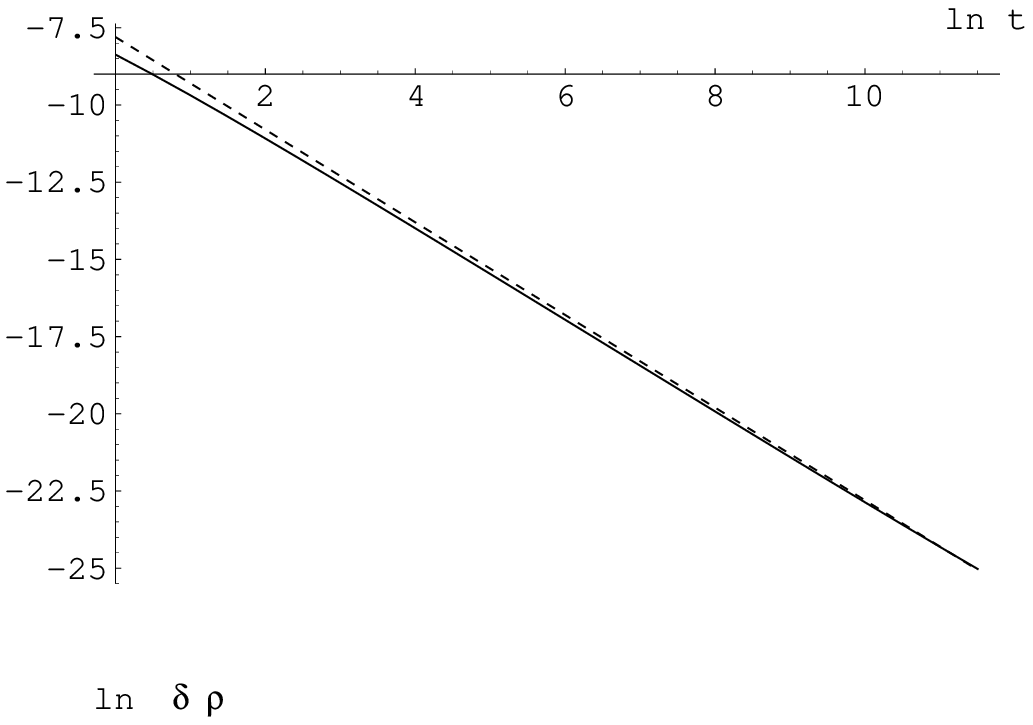,height=37mm,width=70mm} \ 
    
    \vspace{.5in}
    
    \epsfig{file=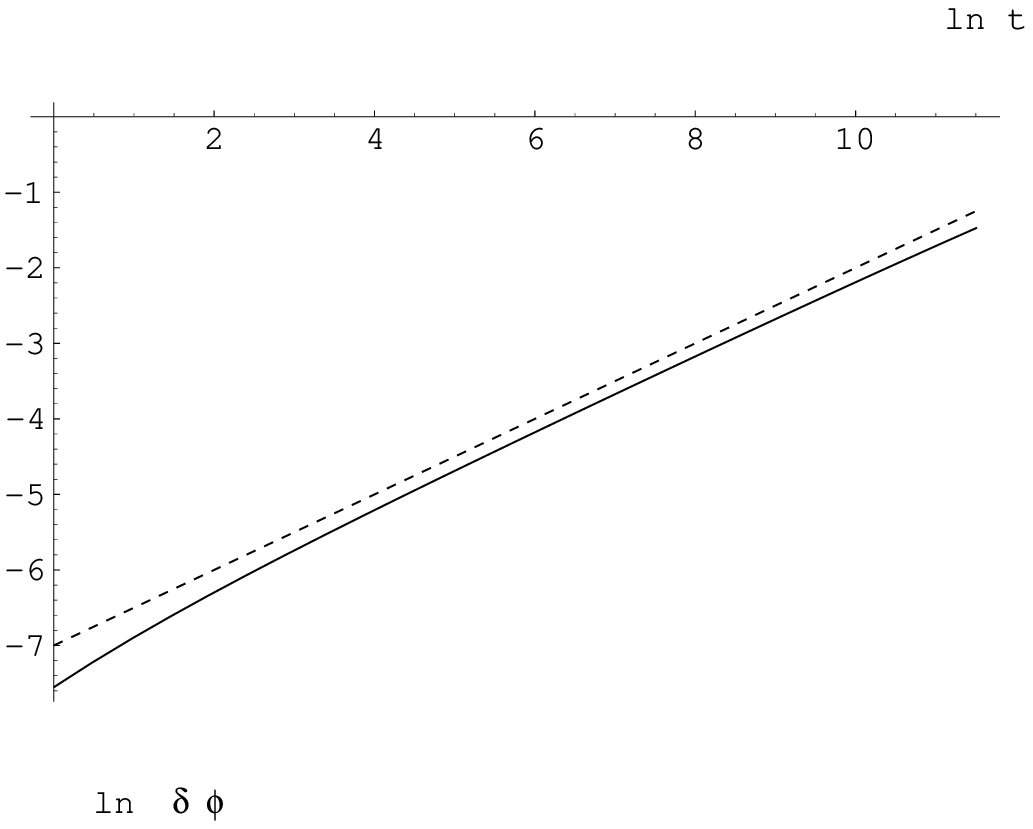,height=37mm,width=70mm}  
  
  \vspace{.5in}
  
    \label{fig:dfeps}
  \end{center}
  
  \caption{ UP: $ \ln \delta H ~{\rm vs.}~\ln t$, 
  dashed=$-1/2$ sloped line. 
             CENTER: $ \ln \delta \rho ~{\rm vs.}~\ln t$, 
             dashed=$-3/2$ sloped line.
             DOWN: $ \ln \delta\phi ~{\rm vs.}~\ln t$,
             dashed=$1/2$ sloped line.}
\end{figure}
}

\begin{figure} 
  \begin{center}
  
   \vspace{.5in}
  
    \epsfig{file=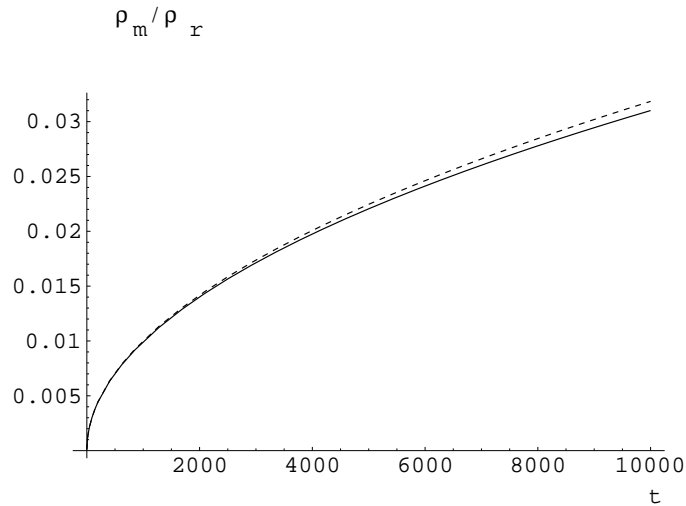,width=90mm}
    
     \vspace{.5in}
    
  \caption{$ \rho_m/\rho_r~{\rm vs.}~t $: solid=with massless $\phi$, 
  dashed=without.}
  \label{fin1.eps}
  \end{center}
\end{figure}

Finally, since perturbation theory is applicable, we conclude
that the presence of a 
massless dilaton does not substantially alter the evolution of 
monopole density in  RD  universe. Figure \ref{fin1.eps} shows 
a numerical solution for $\rho_m/\rho_r $ with and without a 
massless dilaton.

We will now consider the regular mass term, and ``shut off" the
other  part of the potential by putting $\rho_m=0 $. Equation (\ref{dil}) now
takes  the form $\ddot\phi+3H\dot\phi+m^2\phi=0$. We are interested in 
the case of ``slow-roll" in which the friction of the expansion
and the potential   balance each other, and acceleration is approximately zero, 
$ \ddot\phi\approx 0 $. Solving this equation we obtain
\begin{eqnarray}
  & \phi=\phi(0)e^{K(t_0^2-t^2)},~~~K=\frac{m^2}{6H(1)}. & \nonumber
\end{eqnarray}
As long as the initial ratio of $m/H$ is small enough, ensuring  that
the expansion time is shorter than the period of oscillations, the deviation from $
\phi=\phi_0 $ is small. 

\begin{figure} 
  \begin{center}
  
  \vspace{.5in}
  
    \epsfig{file=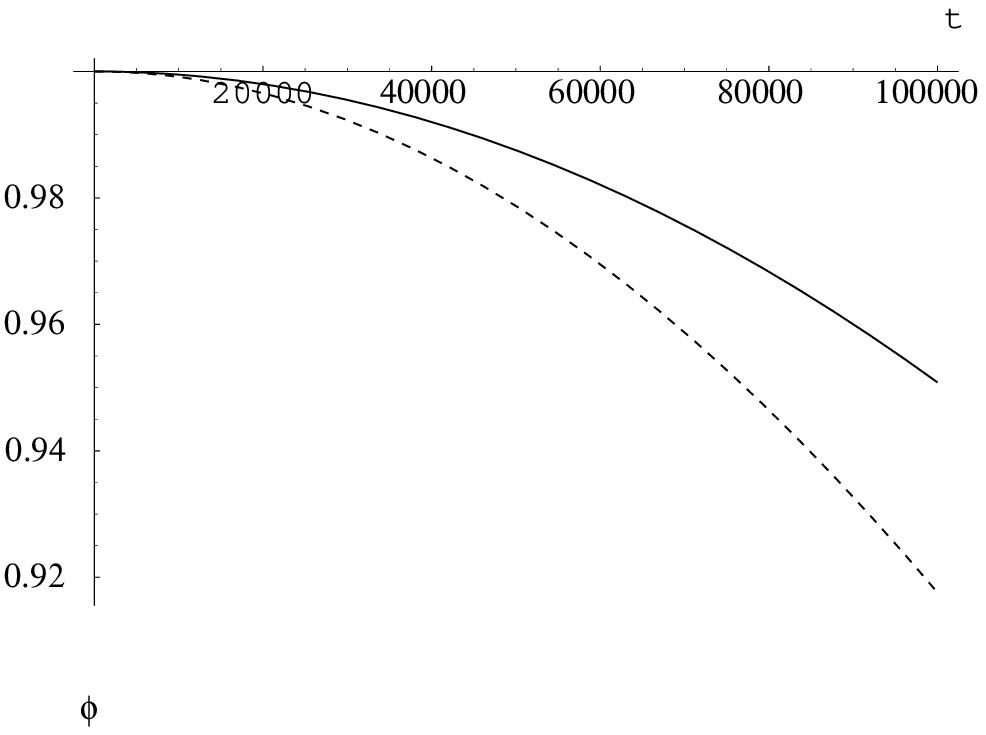,width=80mm}\ \ \ 
    \epsfig{file=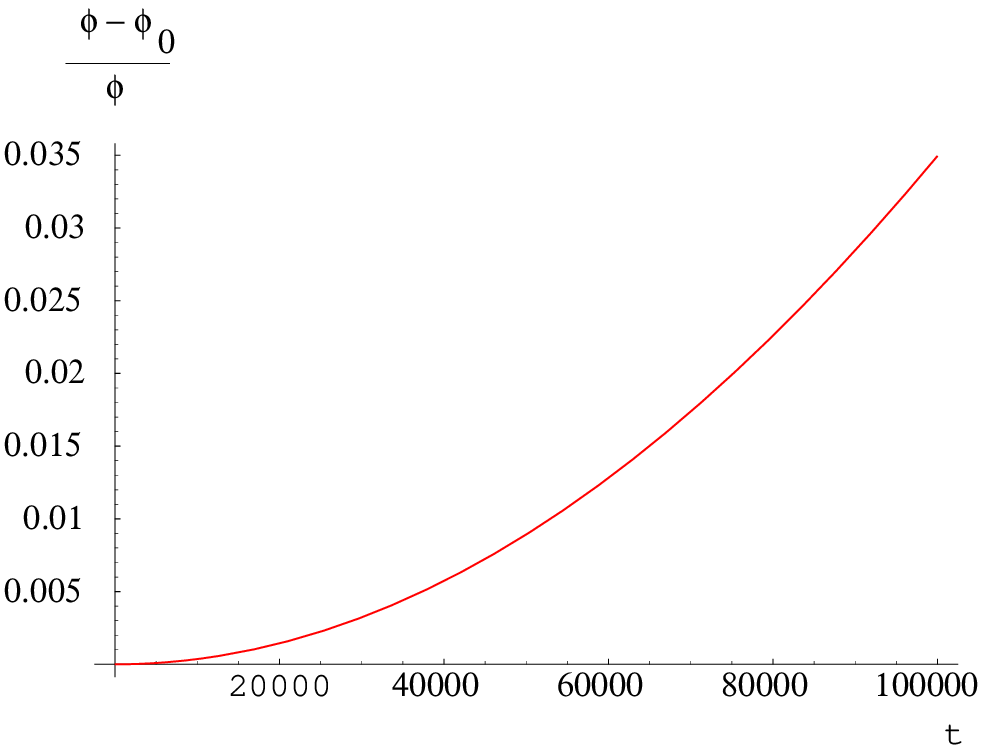,width=80mm}
    
    \vspace{.5in}
    
    \caption{LEFT: $\phi~{\rm vs.}~t$ for a massive dilaton 
    without monopoles:
     solid=numerics, dashed=our estimate.
             RIGHT: Relative error in our estimate.}
    \label{ef2.eps}
  \end{center}
\end{figure}

Figure  \ref{ef2.eps}  shows a numerical solution for 
$\phi$ compared with our estimate, as well as the relative error.

The addition of mass, still keeping the low density condition, produces a 
 small deviation from a constant dilaton, and therefore does not interfere 
with the standard evolution.

\begin{figure} 
  \begin{center}
  
   \vspace{.5in}
   
    \epsfig{file=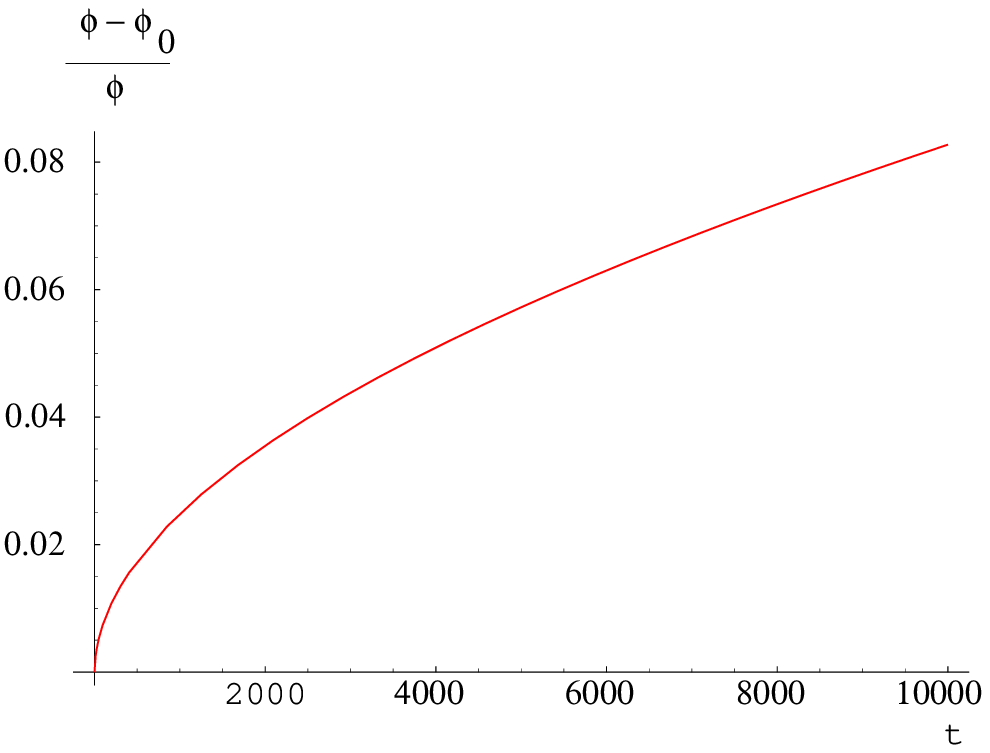,width=80mm} \ \ \ 
    \epsfig{file=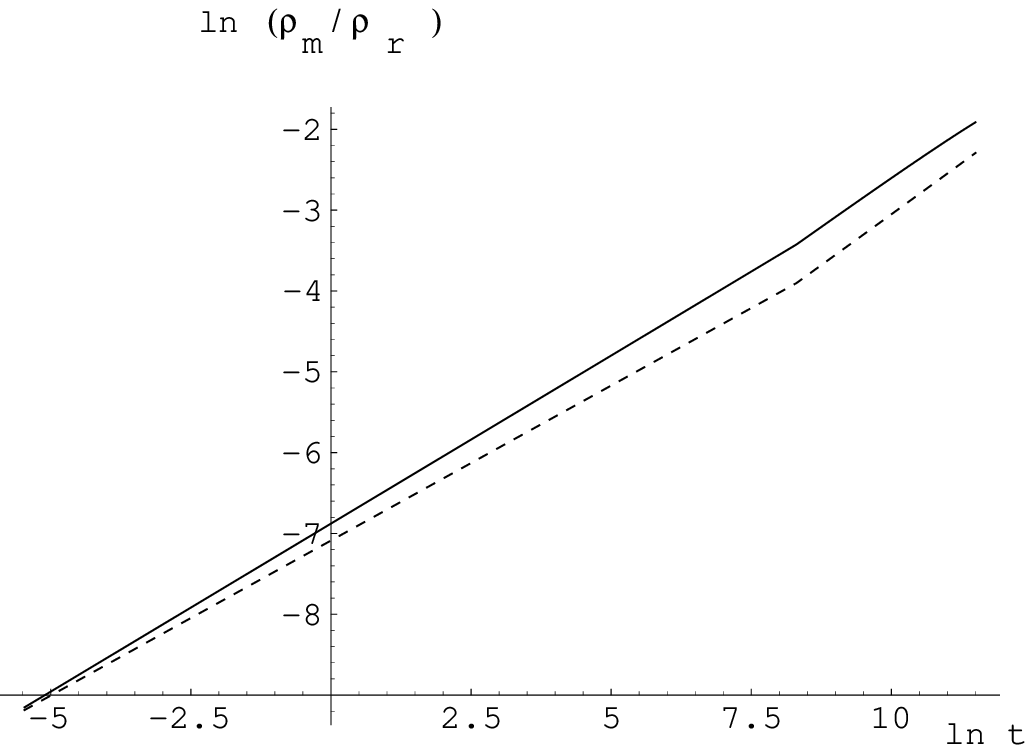,width=80mm}
    
     \vspace{.5in}
     
  \end{center}
  \caption{LEFT: low density massive dilaton: $ \frac{\phi-\phi_0}{\phi} $ in the
                 presence of monopoles.
           RIGHT: $ \ln(\rho_m/\rho_r)~{\rm vs.}~\ln t$: 
           solid=with low density
                  massive $\phi$, dashed=without.}
  \label{fin.eps}
\end{figure}

Now we would like to consider the two potential terms  together. 
Since  the deviation
of the dilaton from a constant in both cases is small, we have no reason to
 believe that the dilaton will act radically different now. The numerical
 solution shows that this is indeed a good assumption. The first graph in Figure
 \ref{fin.eps}  shows the relative error in estimating the massive low density
dilaton in the presence of monopoles as a constant, the second graph shows 
$\rho_m/\rho_r $ with 
 the presence of a low density massive dilaton, compared with the standard
evolution.

\section{High Moduli Density}

High moduli density era is defined to commence when  moduli density 
becomes comparable to the radiation density. 
In this era, there is no
essential difference between the dilaton (assuming it is heavy enough)
 and any of the moduli, since the important element is the oscillations 
 around the minimum of the potential rather
than the coupling to matter. We have checked this assumption
numerically in many cases. Moduli behave here as  NR matter, and the
universe is MD.

The time dependence of the expansion of the universe with and without the moduli is 
different, therefore we use for comparison between the two  not time 
but temperature. Using the facts that during RD 
$ H=1.66\sqrt{g_*}\frac{T^2}{m_p}$ (where $g_*$ is the effective number of
degrees of freedom),
 and that in our model the evolution begins and ends with RD, 
we can use H instead of T. 
We are interested in how the limit on initial ratio of monopole to radiation densities 
differs when  moduli get added to the cosmic mix.

\subsection{A Simple Model}

The basic idea here is that since during oscillations moduli  behave as
NR matter, the ratio of moduli to radiation energy will grow
roughly as the scale factor a(t), assuming that  $g_*$ is constant or slowly
varying during this period. The universe will very quickly
reach  moduli  domination era in which 
$ \rho_{tot}\simeq\rho_{\phi}\propto a^{-3} $, 
and  $\rho_m/\rho_{tot}\propto Const $. 

\vbox{
 \begin{figure} 
  \begin{center}
    
    \vspace{.5in}
    
     \epsfig{file=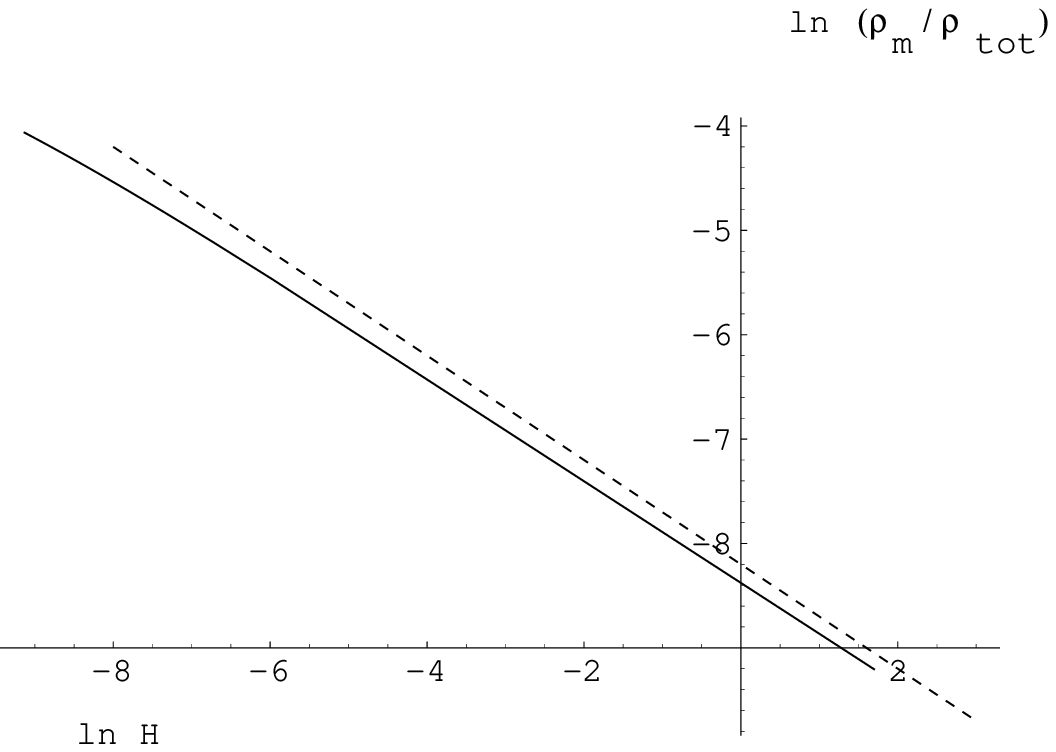,width=80mm}\ \ \ 
    \epsfig{file=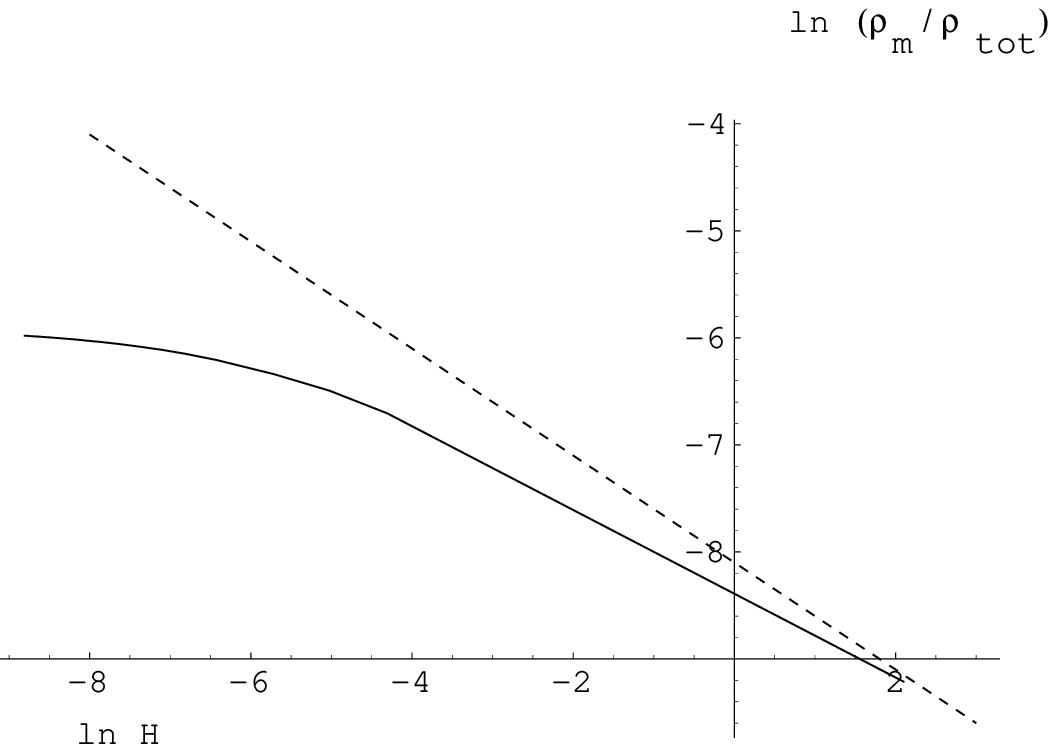,width=80mm}
    
    \vspace{.5in}
    
    \epsfig{file=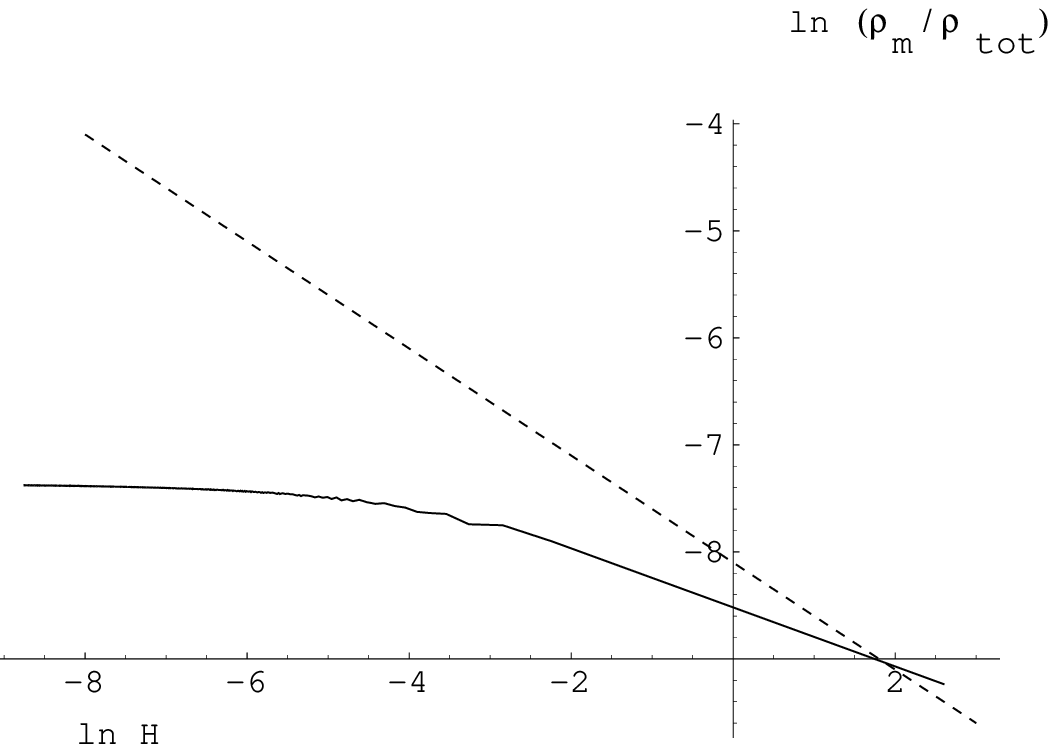,width=80mm} \ \ \
    \epsfig{file=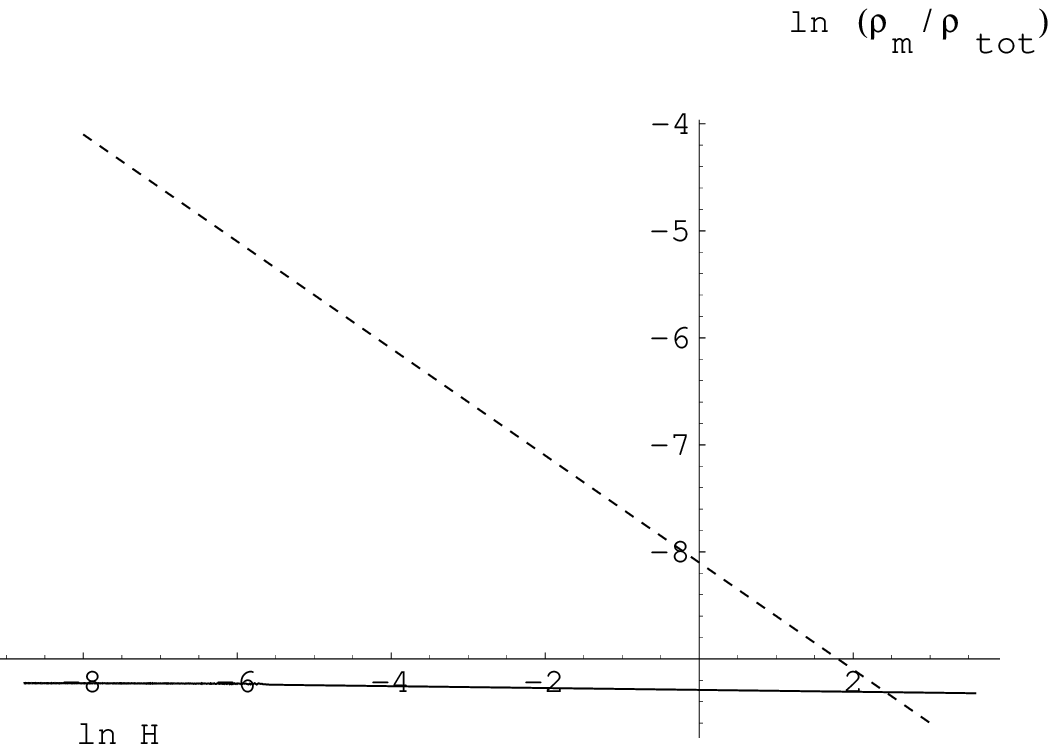,width=80mm}
    
     \vspace{.5in}
    
\caption{$ \ln(\rho_m/\rho_{tot})~{\rm vs.}~\ln H$ for various initial
             conditions. Solid=numerics, dashed$=-1/2$ sloped line.
             UP LEFT: $\rho_m:\rho_r:\rho_{\phi}=1:9999:3 $.
             UP RIGHT: $\rho_m:\rho_r:\rho_{\phi}=1:9999:75 $.
             DOWN LEFT: $\rho_m:\rho_r:\rho_{\phi}=1:9999:300 $.
             DOWN RIGHT: $\rho_m:\rho_r:\rho_{\phi}=1:9999:1200 $.}
    \label{cop4.eps}
  \end{center}
\end{figure}
}

Eventually, moduli get converted  back into radiation and standard cosmology
emerges, with a  diluted $ \rho_m/\rho_r $. In the simplest model
we parametrize the duration of  moduli domination, and
imagine the  conversion into radiation as an instantaneous and completely
efficient process.  The graphs in figure  \ref{cop4.eps}  show a numerical
solution of $\ln(\rho_m / \rho_{tot})~{\rm vs.}~\ln H  $.  As we can see,
a reasonable estimate for $\ln(\rho_m/\rho_{tot})$  is provided by a
$-1/2$ slope for RD, and a horizontal line for MD. At an
arbitrary point $H_f$, the moduli decay instantaneously 
into radiation  and therefore once again  
   
 \begin{eqnarray}
     & \ln(\frac{\rho_m}{\rho_{tot}})\sim \ln(\frac{\rho_m}{\rho_{r}})\sim 
       \ln(a(t))\sim \frac{1}{2}\ln t \sim-\frac{1}{2}\ln H.  & 
  \end{eqnarray}
 
 \begin{figure} 
  \begin{center}
    \epsfig{file=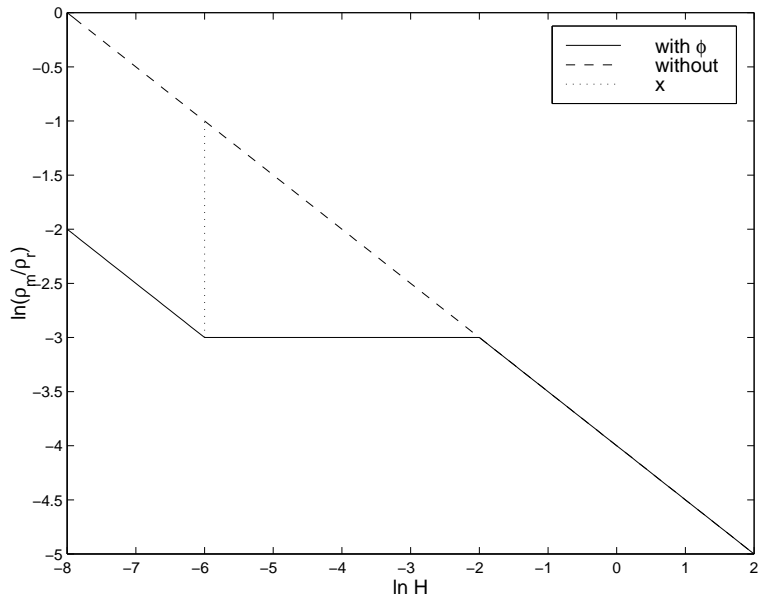,height=80mm}
\caption{$ \ln(\rho_m/\rho_{tot})~{\rm vs.}~\ln H$ with and without the 
         moduli.}
    \label{simple.eps}
  \end{center}
\end{figure}

The difference in the allowed initial monopole to total energy densities 
can be determined by the 
vertical distance between the $-1/2$ sloped line 
and the horizontal line at 
$ H_f$ (see Fig. \ref{simple.eps}). 
A simple geometrical calculation shows that the 
relation between the  duration of the moduli 
domination and the effective amount
of volume expansion efolds it can replace is given by
\begin{equation}
\left(\rho_m/\rho_{tot}\right)_{\rm with}\equiv e^{-x}
      \left(\rho_m/\rho_{tot}\right)_{\rm without},  
      \end{equation}
 where  
\begin{equation}
x=\frac{1}{2}\ln\left(\frac{H_i}{H_f}\right).
\end{equation}

\subsection{A More Realistic Model}

Now we would like to treat decay of moduli to radiation in a better way, 
by including a decay rate in the equations
which will govern the  duration of moduli domination period. 
Dimensional arguments lead to the estimate
\begin{equation}
      \Gamma\equiv\tau^{-1}\sim m^3_{\phi}/ m^2_{pl}. 
\label{gamma}
\end{equation}
The inclusion of such a term is standard \cite{eu}, and leads to   
the approximate equation $\dot\rho_{\phi}+3H\rho_{\phi}=0$, 
from which it is clearly seen that moduli really behave as NR matter. 
This approximate equation
holds as long 
as we can replace $ \phi^2 $ with $ \langle\phi^2\rangle $, 
which is justified 
when  oscillations of  moduli are much faster than the expansion rate. \\
If we want to describe the moduli decaying into photons, or other forms 
of massless particles, we need to 
correct the conservation equations, moduli density decreases
 and the radiation density increases,
\begin{eqnarray}
    & \dot\rho_{\phi}+3H\rho_{\phi}=-\Gamma\rho_{\phi} & \nonumber \\
    & \dot\rho_{r}+4(H+\frac{1}{2}\dot\phi)\rho_{r}=
    \Gamma\rho_{\phi}, & 
\end{eqnarray}
while monopole number conservation still holds. Because the moduli's oscillations 
are very fast and we are using averaged values:
\begin{eqnarray}
    & \langle\phi\rangle=\langle\dot\phi\rangle=0.
\end{eqnarray}
 The approximate equations are given by
\begin{eqnarray}
    & \dot\rho_{\phi}+3H\rho_{\phi}=-\Gamma\rho_{\phi} & \nonumber \\
    & \dot\rho_r+4H\rho_r=\Gamma\rho_{\phi} & \nonumber \\
    & 3H^2=(\frac{1}{4}\dot\phi^2+\frac{1}{4}m^2\phi^2)+\frac{1}{2}
      \rho_r\approx
      \frac{1}{2}\rho_{\phi}+\frac{1}{2}\rho_r, & \nonumber
\end{eqnarray}
where in the last equation we have assumed low monopole density.
Solving for $ \rho_{\phi} $ we obtain
\begin{eqnarray}
     & \rho_{\phi}=
     (\rho_{\phi})_{ 0}e^{-\Gamma(t- t_0)}(\frac{R}{R_0})^{-3}, & 
\end{eqnarray}
where $ t_0 $ is the time when oscillations start.
Solving for $ \rho_r $ we obtain
\begin{eqnarray}
     & \rho_r=\left[(\rho_r)_0-\frac{12}{5}\Gamma H_0\right]\
     \left(\frac{R_0}{R}\right)^4+\frac{12}{5}
       \Gamma H_0(\frac{R_0}{R})^{3/2}. & \nonumber
\end{eqnarray}
Therefore soon after moduli begin dominating, 
radiation density decreases 
only as $ R^{-3/2} $.

\begin{figure} 
  \begin{center}
    \epsfig{file=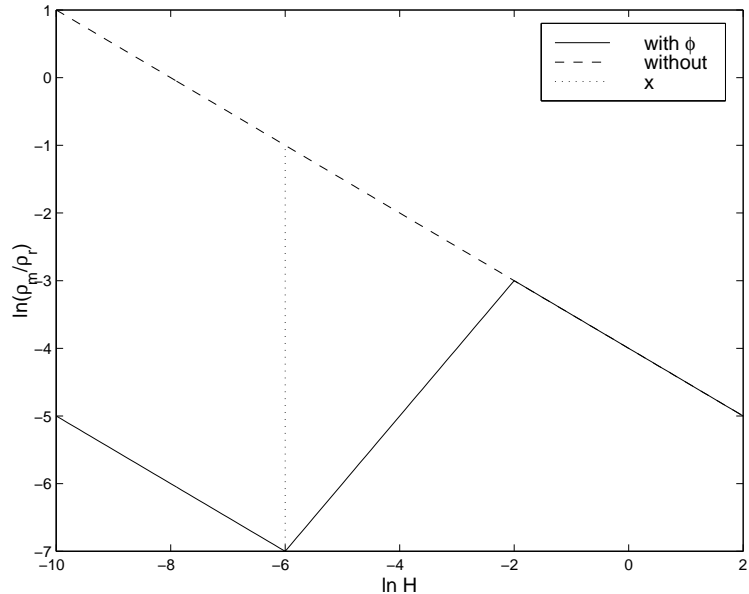,height=80mm}
  \end{center}
  \caption{$ \ln(\rho_m/\rho_{r})~vs.~\ln(H) $: solid=with $\phi$: our
             approximation, dashed=without $\phi$}
  \label{mod.eps}
\end{figure}

As explained, $t_0$ is the time which oscillations begin, which is 
approximately when the moduli come  into domination, 
therefore, $ H_0=\sqrt{\rho_{\phi}}/m_p\approx m_{\phi}^2/m_p $. 
At about $ t_{rh}=\tau_{\phi}=\Gamma^{-1} $, 
most of the moduli have decayed into radiation, and the universe 
re-enters its adiabatic evolution. So between $ H_0=m_{\phi}^2/m_p $ and $ H_{rh}=1/2t_{rh}=\Gamma/2 $, 
the ratio of monopole to radiation density decreases as $ R^{-3/2}\propto H $.  
Figure \ref{mod.eps} shows our approximation for the behavior of 
$\rho_m/\rho_r$ with and without  moduli.
As before, $x$ is the effective amount of volume efolds  moduli replaces 
\begin{eqnarray}
  & (\rho_m/\rho_r)_{with}=
  e^{-x}(\rho_m/\rho_r)_{without} &  \\
  & x=\frac{3}{2}\ln(\frac{H_0}{H_{rh}})=\frac{3}{2}\ln(\frac{2m_{\phi}^2}
  {\Gamma m_p}). & \nonumber
\end{eqnarray}
Using the estimate for $\Gamma$ (\ref{gamma}) gives 
\begin{eqnarray}
      & x=\frac{3}{2}\ln(\frac{2m_p}{m_{\phi}}). & \label{x}
\end{eqnarray}

Figure \ref{real5.eps} shows some comparisons between
our approximation and an exact numerical solution for $\rho_m/\rho_r$. The
scale on the the vertical axis $\ln(\rho_m/\rho_r)$ is determined by the choice
of initial conditions, and $H$ is in our standard units in which $m_p=\sqrt{16\pi}$.

\vbox{
\begin{figure}  
  \begin{center}
    \epsfig{file=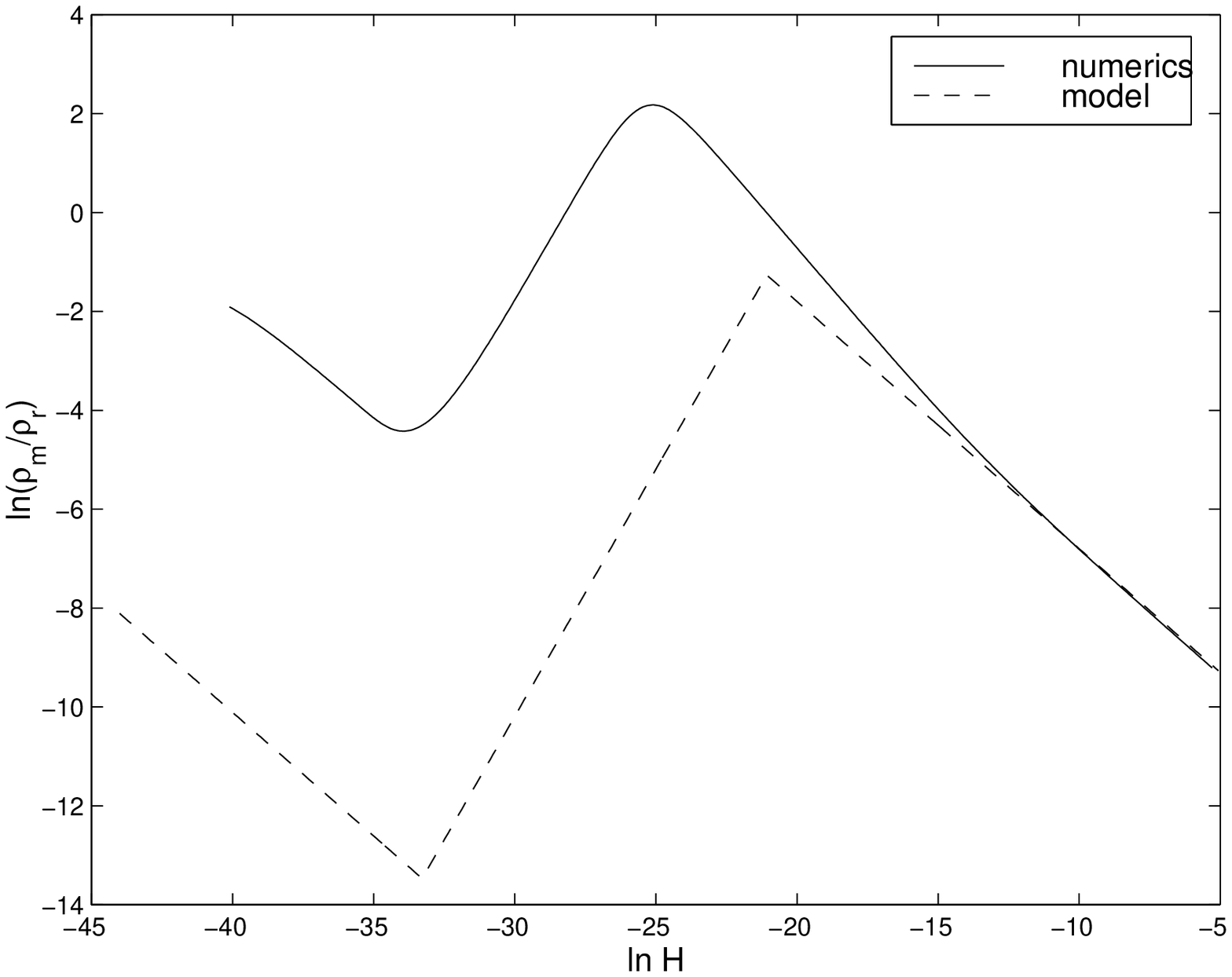,width=80mm}
    \epsfig{file=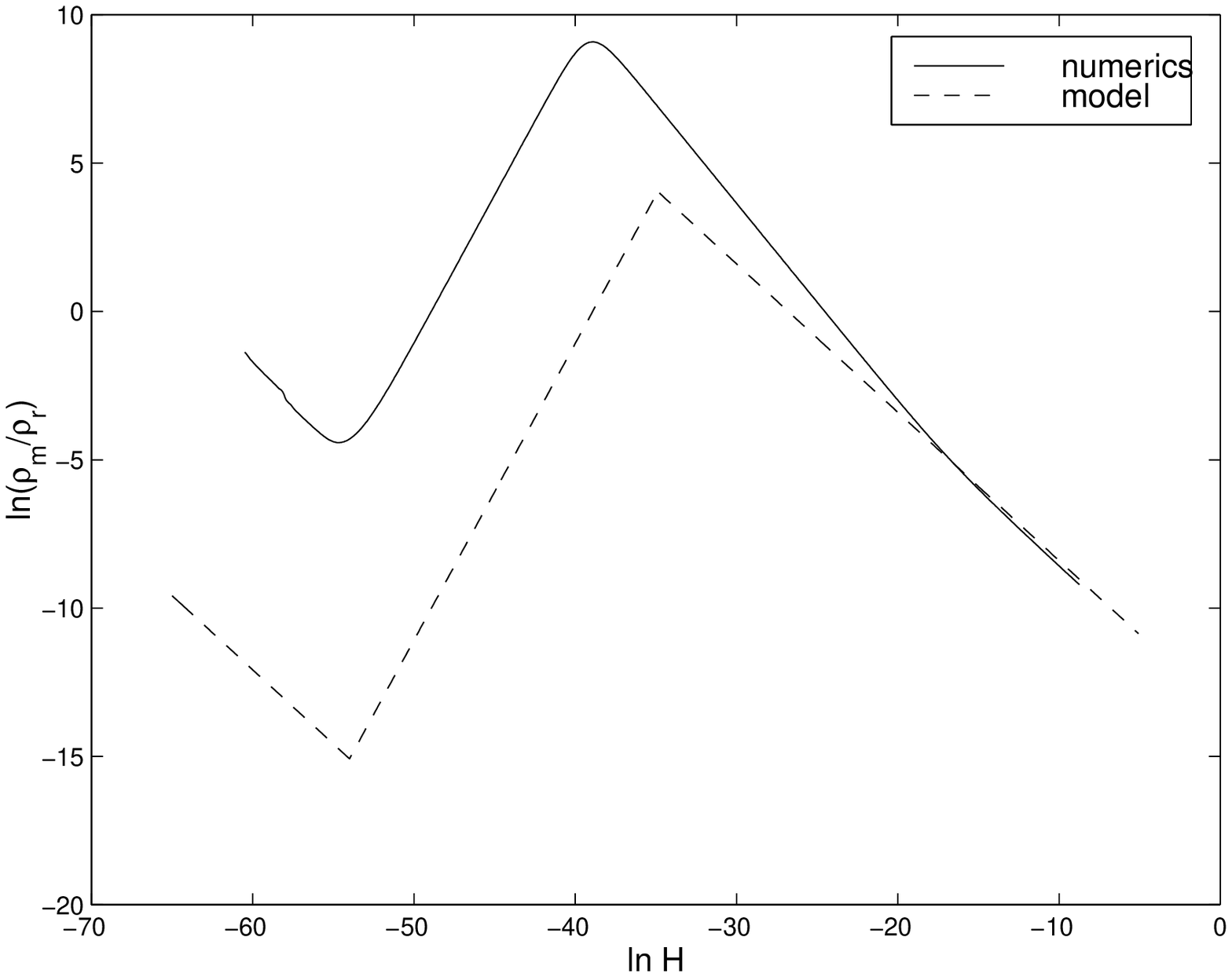,width=80mm}
    
    \epsfig{file=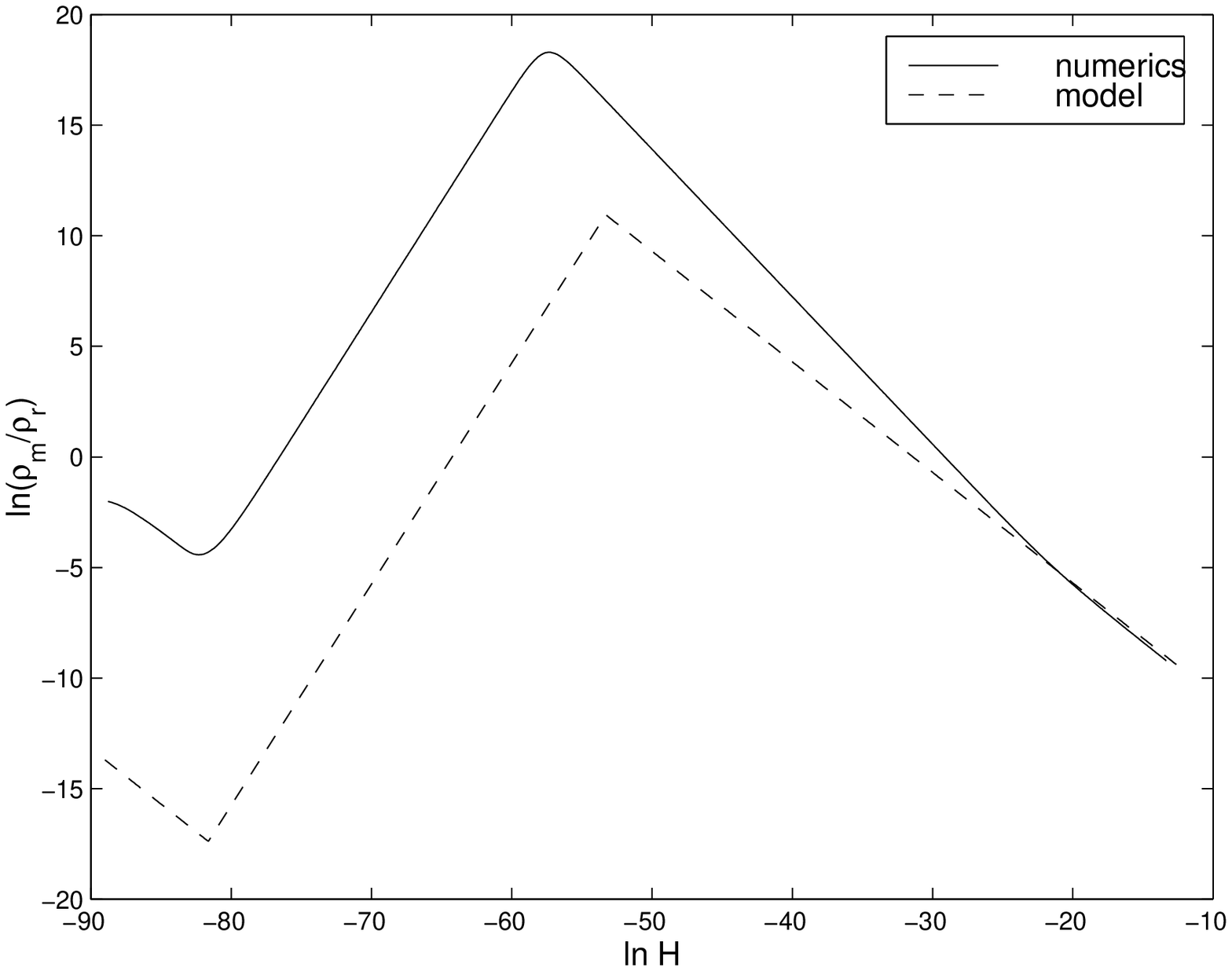,width=80mm}
    \epsfig{file=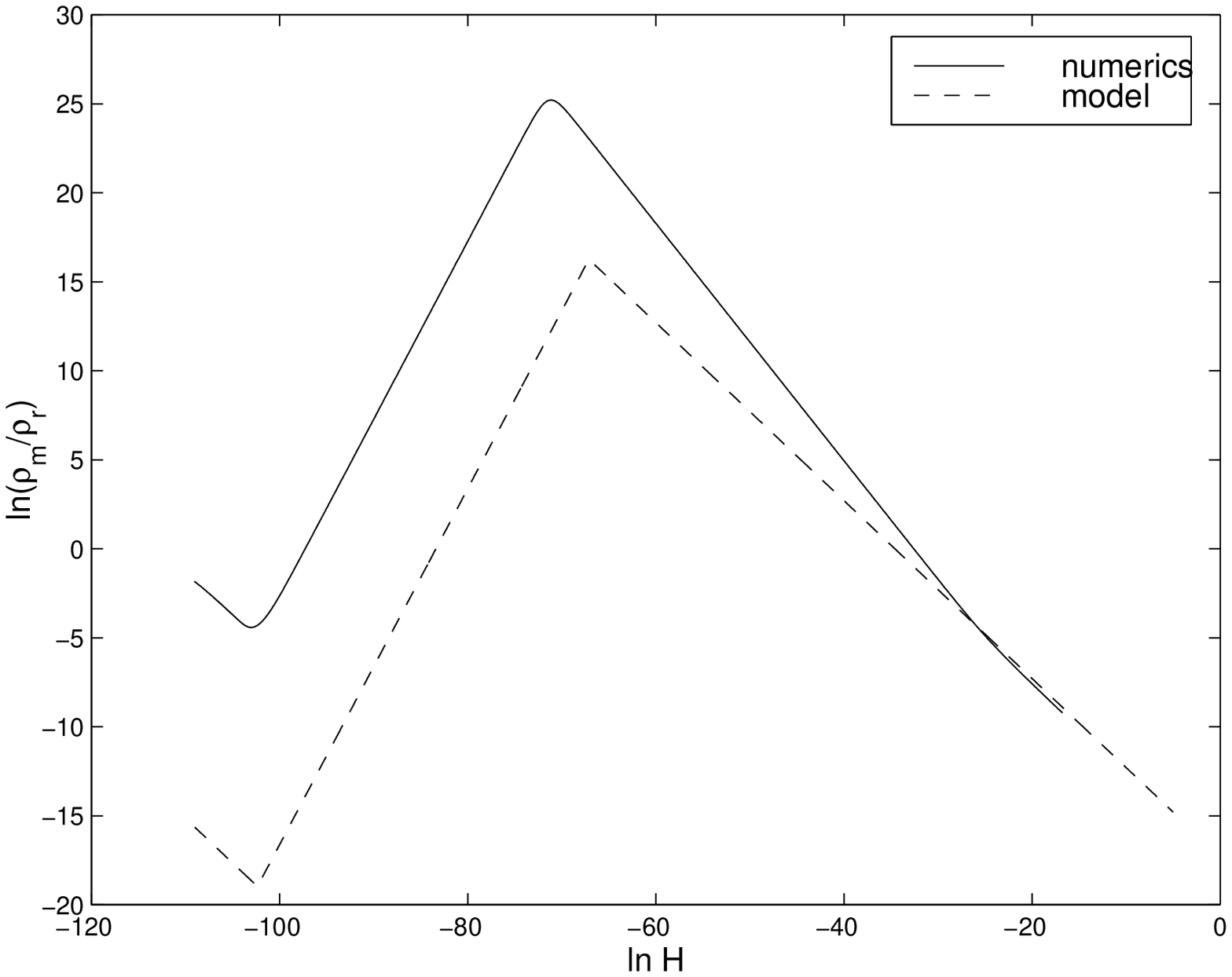,width=80mm}
    
 \caption{$ \ln(\rho_m/\rho_r)~vs.~\ln(H)$, solid=numerics, dashed=our
             model.
             UP LEFT: $ m=10^{-5}m_p,~\Gamma=10^{-15}m_p $,
             UP RIGHT: $ m=10^{-8}m_p,~\Gamma=10^{-24}m_p $,
             DOWN LEFT: $ m=10^{-12}m_p,~\Gamma=10^{-36}m_p $,
             DOWN RIGHT: $ m=10^{-15}m_p,~\Gamma=10^{-45}m_p $.
             }
    \label{real5.eps}
  \end{center}
\end{figure}  
}

\subsection{Moduli relax the monopole bound}
Using (\ref{x}), we want to derive some consequences regarding how moduli 
relax the monopole problem.
\begin{itemize}
  \item
  As was mentioned in the introduction, in order to solve 
  the monopole problem, we  need 27 efolds in volume expansion. 
  If one of the moduli is to supply all 27 on its own, its mass has
  to be below the upper bound 
  \begin{eqnarray}
     & \frac{3}{2}\ln(\frac{2m_p}{m_{\phi}})\simgt 27 & \nonumber \\
     & m_{\phi}\simlt 10^{-8}m_p\approx 10^8~TeV. & 
  \end{eqnarray}
  \item
  Since we do not want the monopoles to interfere with nucleosynthesis, 
  we should demand 
  $ T_{rh}>10~MeV $. We also need to keep in mind that
  baryogenesis has to occur later. The relationship between the moduli mass and the reheat
  temperature is:
  \begin{equation}
      H_{rh}=1.66\sqrt{g_*}\frac{T^2_{rh}}{m_p}
     \sim \frac{\Gamma}{2}=\frac{m_{\phi}^3}{2m_p}, 
   \end{equation}  
   which requires 
\begin{equation}
       m_{\phi}>10^{-14}m_p(\frac{g_*}{100})^{1/6}\approx 100~TeV.   
  \end{equation}
  \item
  Moduli should dilute the monopole 
  to radiation ratio before the universe reaches monopole domination. 
  This means that we need 
  $ \rho_m/\rho_r$ at its highest point, $ H_i $, to be less than
  unity
  \begin{eqnarray}
     & (\rho_m/\rho_r)_i=(\rho_m/\rho_r)_c
     \sqrt{\frac{H_c}{H_i}}<1,
  \end{eqnarray}  
  where   $H_c\approx 10^{10} \sqrt{g_*/100} (T_c/10^{14} GeV)^2 GeV$
  and $(\rho_m/\rho_r)_c\approx 10^{-11} (M_{m}/10^{16} GeV)(10^{14}
  GeV/T_c)$, leading to a lower bound on moduli mass,
   \begin{equation}
    m_{\phi}>3200\left(\frac{g_*}{100}\right)^{1/4}
    \left((\frac{M_m} {10^{14}GeV}\right)GeV. 
  \end{equation}  
 
  \end{itemize}
  
  To summarize our results, here is a table with various values for 
  moduli masses, the amount of (volume) efolds it replaces, 
  its reheat temperature,
  and whether it acts before monopole domination,
   
    \begin{center}
    \vbox{
    \begin{table}
    \begin{tabular}{||c|c|c|c||}
    \hline \hline
    $ \bf{m_{\phi}} $ & \bf{x} & $ \bf{T_{rh}} $ & \bf{before md} \\
    \hline\hline
    $ 10^{-4}~eV $ & $ 112 $ & $ 2\cdot 10^{-30}~MeV $ & no \\
    \hline
    $ 10^5~GeV $ & $ 49 $ & $ 2~MeV $ & yes \\
    \hline
    $ 10^8~GeV $ & $ 39 $ & $ 6\cdot 10^4~MeV $ & yes \\
    \hline
    $ 10^{11}~GeV $ & $ 29 $ & $ 2\cdot 10^9~MeV $ & yes \\
    \hline
    $ 10^{12}~GeV $ & $ 25 $ & $ 6\cdot 10^{10}~MeV $ & yes \\
    \hline
    $ 10^{14}~GeV $ & $ 18 $ & $ 6\cdot 10^{13}~MeV $ & yes \\
    [.3ex] \hline\hline
  \end{tabular}
  \caption{$x$ values for different moduli mass.}
  \label{tb5}
  \end{table}
  }
  \end{center}
 
\section{Discussion}

\begin{itemize}

\item
  The mechanism described here makes use of  massive scalar fields 
  that decay 
  into radiation. String theory provides us with several candidate fields 
  -- moduli (including the dilaton). Instead of demanding that a 
  single field  provides all the 27 efolds, a possible scenario is 
  that two or maybe more fields have different masses and 
  therefore oscillate at different times, and each 
  of them contributes a few of the needed efolds. As an example, consider 
  two fields, one with mass of $ 10^8~GeV $, and the other with mass of 
  $ 10^{12}~GeV $. We will assume that for both fields the decay rate is
  given by (\ref{gamma}). The $ 10^{12}~GeV $ mass field begins 
  to dominate when $ \ln(H_i)=\ln(\frac{m^2}{m_p})\sim 11.5 $, 
  and ends at $\ln(H_{rh})=\ln(\frac{m^3} {m^2_p})\sim-5.2 $. 
  During this time it provides 25 efolds. Only later does
  the $ 10^8~GeV $ mass field reaches  domination, at $ \ln(H_i)=-6.9 $ until 
  $ \ln(H_{rh})\sim -32.9 $, and it provides additional 39 efolds. 
  Therefore, it is possible to use
  use several heavier fields instead of a single lighter one.

\item

  The presence of moduli changes the standard adiabatic evolution 
  of the universe, 
  introducing a period (or
  periods) of matter domination. Looking at how many decades of temperature
  were indeed matter dominated will quantify the deviation from the standard
  evolution,
  \begin{eqnarray}
     & \ln(\frac{T_i}{T_{rh}})=\ln(\sqrt{\frac{H_i}{H_{rh}}})\sim
     \frac{1}{2}  \ln(\frac{\Gamma m_p}{2m^2}), & 
  \end{eqnarray}
  and using (\ref{gamma}),
  \begin{eqnarray}
     &  \ln(\frac{T_i}{T_{rh}})\sim\frac{1}{2}\ln(\frac{m}{2m_p}). & 
  \end{eqnarray}
  Looking at (\ref{x}) shows that the relation between $x$ and 
  the deviation is
  linear. We conclude that to substantially relax the monopole bound
  long periods of coherent oscillation are required. 
\end{itemize}

  We want to to compare our estimate to the numerical results. 
  Table  \ref{tb1}  
  shows the numerical and the estimated $x$ values, and the relative error 
  ($ E_x=\frac{x_{num}-x_{est}}{x_{num}} $):
    
    \begin{center}
    \vbox{
    \begin{table}
    \begin{tabular}{||c|c|c|c|c||}
    \hline \hline
    $ \bf{m} $ & $ \bf{\Gamma} $ & $\bf{x_{est}}$ & $\bf{x_{num}} $ &
      $\bf{E_x}$ \\
    \hline\hline
    $10^{-5}m_p$ & $ 10^{-15}m_p $ & $ 18.3 $ & $ 9.8 $ & $ -87\% $ \\
    \hline
    $10^{-8}m_p$ & $ 10^{-24}m_p $ & $ 28.7 $ & $ 18.6 $ & $ -54\% $ \\
    \hline
    $10^{-12}m_p$ & $ 10^{-36}m_p $ & $ 42.5 $ & $ 30.3 $ & $ -40\% $ \\
    \hline
    $10^{-15}m_p$ & $ 10^{-45}m_p $ & $ 52.8 $ & $ 39.2 $ & $ -35\% $ \\
    \hline
    $10^{-18}m_p$ & $ 10^{-54}m_p $ & $ 63.2 $ & $ 47.5 $ & $ -33\% $ \\
    [.3ex] \hline\hline
    \end{tabular}
    \caption{Accuracy of analytical estimates of $x$.}
    \label{tb1}
    \end{table}
    }
    \end{center}

  As can be seen, our estimated values improve as the decay rate 
  (and moduli's mass) get smaller. 
  Also, there is a systematic ``overshooting" 
  (estimating too big an $x$).  
  We understand this effect the following way, our calculation assumes 
  that immediately as the moduli come to domination they begin
  to  oscillate 
  and the $ \Gamma $ term becomes effective. Looking at Figure 
  \ref{real5.eps}, we see that there is a period in which  
  moduli dominate  and therefore the universe is MD 
  ($ R\propto t^{2/3} $), yet 
  the $ \Gamma $ term is not operative. This results in 
  $ \frac{\rho_m}{\rho_r}\propto H^{-2/3} $ instead of $ H^{-1/2} $, thus the 
  numerical results give values that are lower than our estimate. 
  
  We can try to improve our simple estimate by taking into account 
  this rise, 
  define $H_m$ such that between $H_i$ and  $H_m$ the universe is MD, 
  but the moduli do not yet oscillate. Only below $H_m$ the oscillations 
  and the moduli  decay start. Following \cite{eu}, 
  $ H_m=\sqrt{\frac{m^2\Gamma}{m_p}} $. Such a calculation will yield 
  $ x=\frac{2}{3}\ln(\frac{2^{9/4}m^2}{\Gamma m_p})=\frac{2}{3}\ln(\frac{2^{9/4}m_p}{m}) $, 
  which compared with the numerical values of $x$ is consistently 
  ``undershooting" 
 (estimating too small an $x$). We have obtained an upper bound on $x$
  as well as a 
  lower bound, and for the best estimated value we can take their average, 
  $ x_{av}=\ln(\frac{2^{3/2}m_p^{13/12}}{m^{13/12}}) $. Table  III
  summarizes the relative error in the upper and lower bounds, as well as in 
  the average,
    \begin{center}
    \vbox{
    \begin{table}
    \begin{tabular}{||c|c|c|c|c|c||}
    \hline \hline
    $\bf{m}$ & $ \bf{\Gamma} $ & $\bf{ x_{av} }$ & $\bf{E_x(upper~bound)}$ &
          $\bf{E_x(lower~bound)} $ & $\bf{E_x(average)}$ \\
    \hline\hline
    $10^{-5}m_p$ & $ 10^{-15}m_p $ & $13.5$ & $ -87\% $ & $ +11\% $ & $ -38\% $ \\
    \hline
    $10^{-8}m_p$ & $ 10^{-24}m_p $ & $21.0$ &$ -54\% $ & $ +28\% $ & $ -13\% $ \\
    \hline
    $10^{-12}m_p$ & $ 10^{-36}m_p $ & $31.0$ &$ -40\% $ & $ +36\% $ & $ -2\% $ \\
    \hline
    $10^{-15}m_p$ & $ 10^{-45}m_p $ & $38.5$ &$ -35\% $ & $ +39\% $ & $ +2\% $  \\
    \hline
    $10^{-18}m_p$ & $ 10^{-54}m_p $ & $46.0$ & $ -33\% $ & $ +40\% $ & $ +3\% $  \\
    [.3ex] \hline\hline
    \end{tabular}
    \label{tb2}
    \caption{Errors in analytical estimates of $x$.}
    \end{table}
    }
    \end{center}
  To illustrate the two bounds, Figure \ref{double} shows the two bounds 
  and the numerical behavior, for $ m=10^{-15}m_p $.

\vbox{
\begin{figure} 
  \begin{center}
    \epsfig{file=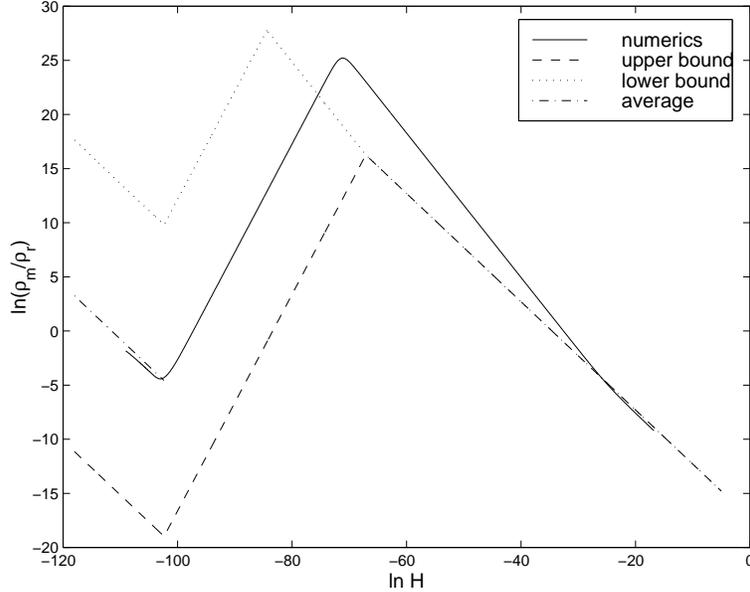,height=80mm}  
   \caption{Approximate evolution, 
   $\Gamma=10^{-15}m_p$ solid=numerics, dotted=upper bound,
                         dashed=lower bound, dashed-dotted=the average.}
    \label{double}
  \end{center}
\end{figure}  
}

\end{document}